\documentclass[12pt]{article}
\usepackage{epsfig}

\def\beq{\begin{equation}}
\def\eeq#1{\label{#1}\end{equation}}
\def\eeqn{\end{equation}}

\def\beqa{\begin{eqnarray}}
\def\eeqa#1{\label{#1}\end{eqnarray}}
\def\eeqan{\end{eqnarray}}

\let\bar=\overbar

\def\Dslash{\not{\hbox{\kern-4pt $D$}}}
\def\dslash{\not{\hbox{\kern-2pt $\del$}}}

\def\msb{{\bar{\ssstyle M \kern -1pt S}}}

\def\Title#1{\begin{center} {\Large {\bf #1} } \end{center}}

\begin{document}

\Title{Measurement of the flavour-specific {\it CP} violating asymmetry $a_{sl}^s$ in $\bar{B}_{s}^0$ decay}

\bigskip\bigskip


\begin{raggedright}  

{\it Zhou Xing\index{}\\
Department of Physics\\
Syracuse University \\
Syracuse, NY, USA \\
Email: zhxing@syr.edu\\
Proceedings of CKM 2012, the 7th International Workshop on the CKM Unitarity Triangle, University of Cincinnati, USA, 28 September - 2 October 2012}
\bigskip\bigskip
\end{raggedright}

\begin{abstract}

The CP violating asymmetry $a_{sl}^s$ is studied with a sample of $\bar{B}_s^0$ or $B_s^0$ semi-muonic decays  in proton-proton collisions at a centre-of-mass energy of 7 TeV at LHCb with an integrated luminosity of 1 fb$^{-1}$.  The final state studied is $D_s^{\pm} \mu^{\mp}$, with $D_s^{\pm}$ reconstructed in the final state $\phi\pi^{\pm}$. The $D_s^{\pm} \mu^{\mp}$ yields are summed over untagged $\bar{B}_s^0$ and $B_s^0$ initial states, and integrated with respect to decay time. Data driven methods have been developed to measure all the efficiency ratios needed to determine $a_{sl}^s$ from the measured raw asymmetry. We obtain $a_{sl}^s$= (-0.24$\pm$0.54$\pm$0.33)\%, where the first uncertainty is statistical and the second systematic. 
\end{abstract}

\section{Introduction}

The goal of this study is the determination of CP asymmetry in $\bar{B}_s^0$--$B_s^0$ mixing, which is a sensitive probe of new physics. In the neutral $B_{s}$ mixing system the time dependent mass eigenstates are related to the weak eigenstates by a 2x2 complex matrix \cite{Nierste:2009wg}. The measurable quantities are the mass difference $\Delta M=M_H-M_L$, the width difference $\Delta \Gamma=\Gamma_L-\Gamma_H$ and the semileptonic  (flavour-specific) asymmetry $a^s_{{sl}}$. Here a small asymmetry term $a^s_{sl}$ is defined and related to the mixing coefficients $q$ and $p$:
\begin{eqnarray}
 a^s_{{sl}} &=&1 - \left| \frac{q}{p} \right|^2. \label{defa} 
\end{eqnarray}

In the Standard Model $a_{sl}^s$ is tiny and beyond current measurement precision. The D0 experiment, however, measured a $3.9$ standard deviation excess for a combination of $a_{sl}$ for $B^0$ and $B^0_s$ mesons (dimuon asymmetry) that is ascribed mostly to the $B_s^0$. Extracting the direct values for $a_{sl}^s$ and $a_{sl}^d$ gives more independent insights. It was realized LHCb could, in principle, check this measurement, but it would be difficult.

For a hadron collider experiment, the initial production asymmetry is not necessarily zero and has to be considered in the derivation of $a^s_{sl}$. To the first order we have, after integrating over time, the untagged asymmetry $A_{meas}$:
\begin{equation}
\label{Eq:acceptrat}
A_{meas}=\frac{\Gamma[D_s^- \mu^+ ]-\Gamma[D_s^+ \mu^-]}{\Gamma[D_s^- \mu^+]+\Gamma[D_s^+ \mu^-]}
=\frac{a^s_{sl}}{2}+\left[a_p-\frac{a^s_{sl}}{2}\right]\frac{\int_{t=0}^{\infty} {e^{-\Gamma t}\cos ( \delta m \, t )\epsilon(t)dt}}
{\int_{t=0}^{\infty} {e^{-\Gamma t}\cosh \frac{\Delta\Gamma \, t}{2}\epsilon(t)dt}},
\end{equation} where the $\epsilon(t)$ is the detector acceptance function and the physical quantity $a_{sl}^s$ is a factor of 2 larger than the measured untagged asymmetry $A_{meas}$.

In principle, we have to be concerned with particle anti-particle production asymmetries, denoted as $a_p$ as well as detector related asymmetries, $a_d$. For this time integrated measurement of $a_{sl}^s$, one key element is based on the realization that the rapid  oscillations cause any production asymmetry between $B_{s}^{0}$-$\bar{B}_{s}^{0}$ to be diluted to a negligible level. Using $\epsilon(t)$ from MC, we have estimated integral ratio in Eq.~\ref{Eq:acceptrat} to be $0.2\%$ for $B^{0}_{s}$ decays and $33\%$ for $B^{0}$ decays. Since the initial B production asymmetry is at most only a few percent, this reduces the effect of $a_p$ to the level of a few $\times 10^{-4}$ for $B_s^0$ decays, well under our goal of the uncertainty on the order of $10^{-3}$.

\section{Analysis Method}

Our goal is to measure the difference between $D_s^+X\mu^-\overline{\nu}$ and $D_s^-X\mu^+\nu$, where the $D_s^{\pm}\to K^+K^-\pi^{\pm}$. In the first measurement we restrict ourselves to $D_s^+$  decaying into $\phi\pi^+$, in order to reduce the effects of charge asymmetries induced by different kinematics of the $K^+$ and $K^-$ in the final state as well as suppress false $D_s^+$ background.

We  construct the measured asymmetry $A_{meas}$ as
\begin{equation}
A_{meas} = \frac{\frac{N(D_s^{-} \mu^{+})}{\epsilon(\mu^{+})}-\frac{N(D_s^{+} \mu^{-})}{\epsilon(\mu^{-})}}{\frac{N(D_s^{-} \mu^{+})}{\epsilon(\mu^{+})}+\frac{N(D_s^{+} \mu^{-})}{\epsilon(\mu^{-})}}
\label{eq:intasy}
\end{equation}
where $N(D_s^{\pm} \mu^{\mp})$ is the measured yield of $D_s^{\pm} \mu^{\mp}$ pairs and $\epsilon(\mu^{+})$ is the efficiency for muon identification and trigger efficiency.

\begin{figure}[hbt]
\begin{center}
\includegraphics[width=5.9 in]{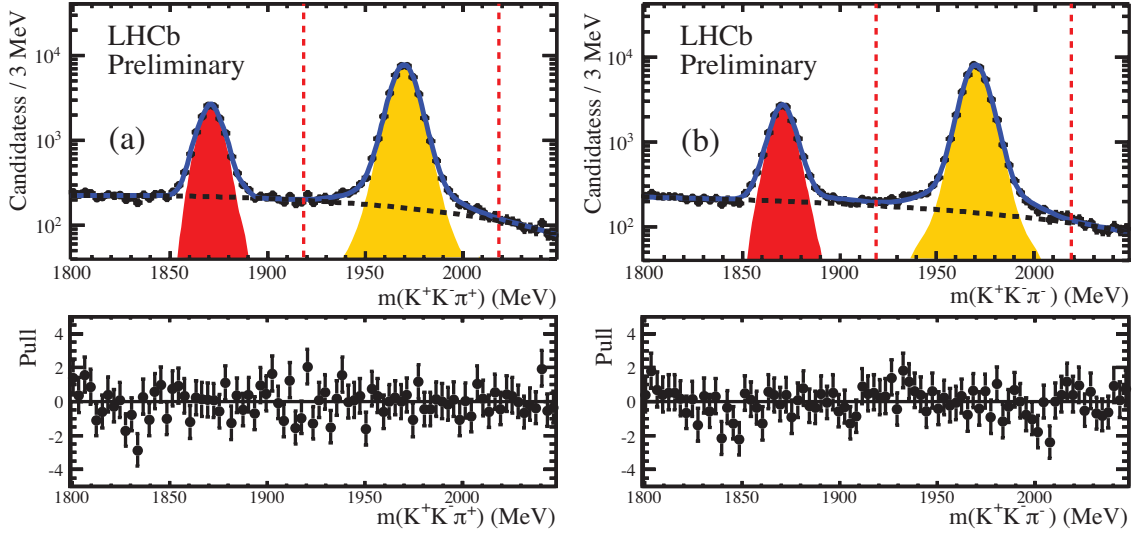}
\end{center}
\vspace{0pt}
\caption{The invariant mass distributions for: (a) $K^+K^-\pi^+$ events and (b) $K^+K^-\pi^-$ events (only the magnet up data is shown here) with m($K^+K^-$) within 20 MeV of $\phi$ meson mass. The fitting functions are described in the text.} 
\label{fig:Signal_mass_fit_up_after_fix}
\end{figure}

We use both fitting and counting (count the number of total events in mass range [1919, 2018] MeV and then subtract off the background level that is obtained from the full fit within the same window) methods to exact the signal yields, which are shown in Fig.~\ref{fig:Signal_mass_fit_up_after_fix}. Here both the signal $D_s^{\pm}$ (yellow shaded area) and $D^{\pm}$ (red shaded area) are fitted by triple Gaussian functions with two Gaussian's share a common mean. The background (black dashed line) is modeled by a second order polynomial function. 

An elegant data-driven approach is developed to measure the pion tracking efficiency thus the relative tracking efficiency between $\pi^{+}$ and $\pi^{-}$ \cite{Dsprod}. The idea is to first ``partially" reconstruct the prompt $D^{*+}\rightarrow D^{0}\pi^{+}_{s}$ decays where $D^{0}\rightarrow K^{-}\pi^{+}\pi^{-}\pi^{+}$ with one $D^{0}$ daughter pion ignored (called partial reconstruction) and subsequently ``fully" reconstruct the whole decay sequence on top of the partial reconstruction (called full reconstruction). The detection efficiency is then determined by the ratio of the full reconstruction to the partial reconstruction. The pion detection asymmetry is then examined versus both momentum and transverse momentum, and no dependence is observed. This fact simplifies calibration of the tracking asymmetry for the charge symmetric final state ($\pi^{\pm}$--$\mu^{\mp}$ pairing). This technique is also used to measure the $D_s^+$ -- $D_s^-$ production asymmetry \cite{Dsprod}.

To determine the muon identification and trigger efficiencies, we developed a minimal biased method to select a large sample of $J/\Psi \rightarrow \mu^{+} \mu^{-}$ events. The idea is to take fully reconstructed hadronic events such as $B \rightarrow D \pi$ that are triggered on the B hadrons, and use $J/\Psi \rightarrow \mu^+\mu^-$ decays in the same event arising from the decay of the companion b quark to give a unbiased measurement of muon efficiency.

\section{Results}

We form an arithmetic average of the magnet up and magnet down data. Taking the average in this manner tends to cancel any residual magnet field related biases. After correcting for muon identification and trigger asymmetry, $a_{d}(\mu)$, the untagged asymmetry between $D_s^{-}\mu^{+}$ and $D_s^{+}\mu^{-}$ final states is measured as $(-0.12\pm0.27\pm0.16)\%$ with the first error statistical and second systematic. Multiplied by a factor of two as shown in Eq.~\ref{Eq:acceptrat}, the physical quantity $a_{sl}^s = (-0.24\pm0.54\pm0.33)\%$ \cite{asl}.

The predictions in the Standard Model for semileptonic asymmetries in $B_s^0$ and $B_d^0$ decays are $a_{sl}^s = (1.9 \pm 0.3) \times 10^{-5}$, and $a_{sl}^d = (-4.1 \pm 0.6) \times 10^{-4}$ \cite{ALenz07}. Our measurement is consistent with the SM prediction.

\begin{figure}[hbt]
\begin{center}
\includegraphics[width=5.9 in]{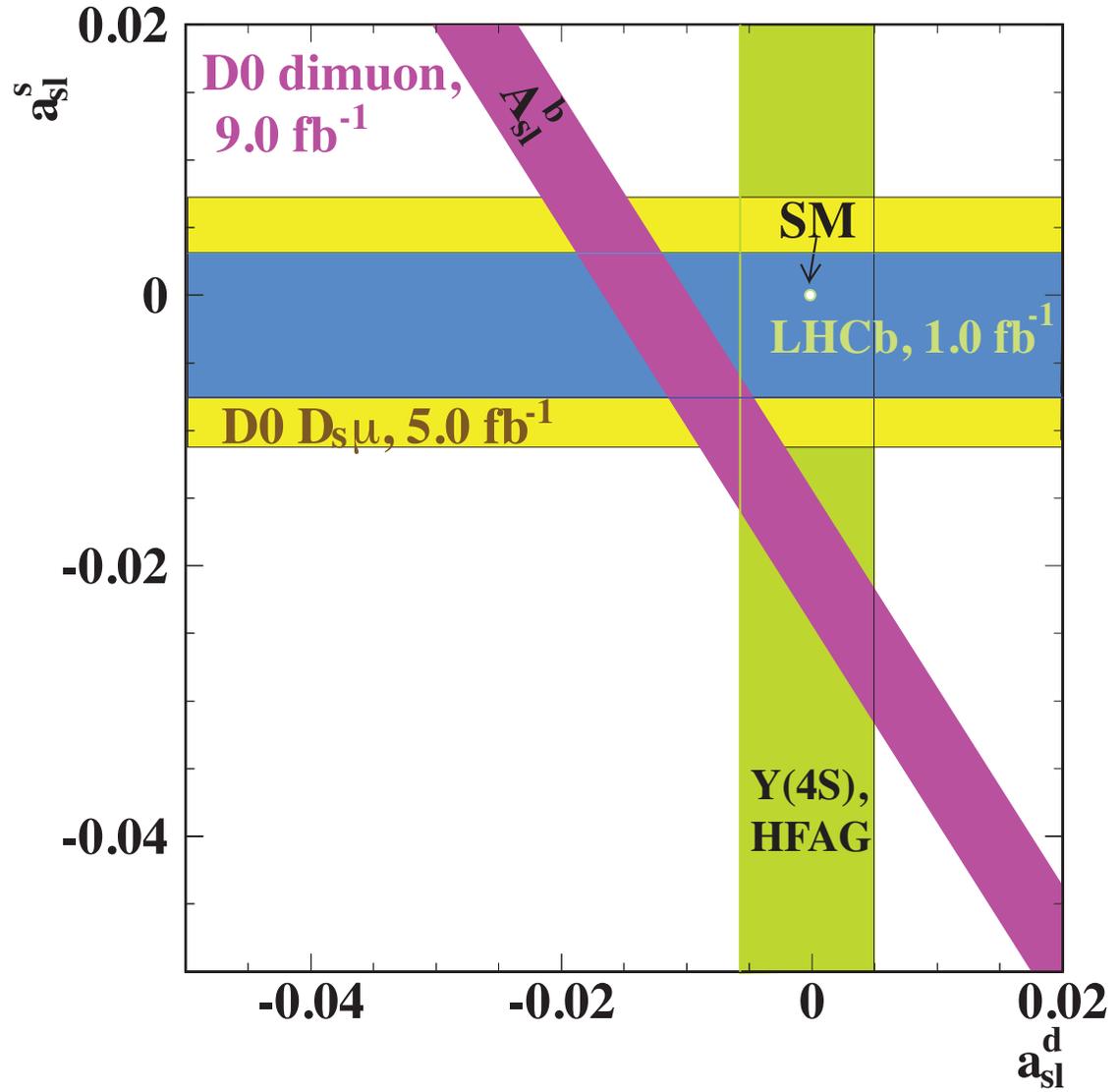}
\end{center}
\vspace{0pt}
\caption{Measurements of semileptonic decay asymmetries. The bands correspond to the central values $\pm1$ standard deviation, defined as the sum in quadrature of the statistical and systematic errors..} 
\label{fig:result}
\end{figure}

We show in Fig.~\ref{fig:result} our measurement, the D0 dimuon result, the previous D0 measurement using flavour tagged $D_s^{\mp}\mu^{\pm}$ events in a 5 ${\rm fb}^{-1}$ sample \cite{D0}, that gives a value of $a_{sl}^s = (-0.17 \pm 0.91^{+0.14}_{-0.15})\%$ \footnote{An updated measurement performed by D0 which supersedes this one results in $a_{sl}^s =(-1.08\pm0.72\pm0.17)\%$. In combination with the D0 dimuon asymmetry this gives a combined result of  $a_{sl}^s =(-1.70\pm0.56)\%$.}, and the average value of $a_{sl}^d$ from $\Upsilon$(4S) measurements of $(-0.05 \pm 0.56)\%$.

In conclusion, our result $a_{sl}^s$ is the most precise determination to date, and is in agreement with the Standard Model prediction and D0 results.

\section*{Acknowledgement}
Work supported by U.S. National Science Foundation.

\end{document}